\documentclass{article}

\usepackage[T1]{fontenc}
\usepackage{url,hyperref,lineno,microtype,subcaption}
\usepackage[onehalfspacing]{setspace}
\usepackage{xspace}

\usepackage{geometry}           
\usepackage{palatino}           
\usepackage[english]{babel}    
\usepackage{graphicx,caption}           
\usepackage[dvipsnames]{xcolor}
\usepackage{indentfirst}        
\usepackage{amsmath,amsfonts,amssymb}
\usepackage{color}
\usepackage{ulem}
\usepackage{url}
\usepackage{bm}
\usepackage{bbm}
\usepackage{float}
\usepackage{mathtools}
\usepackage[toc,page]{appendix}
\usepackage{setspace}
\usepackage{xpatch}
\usepackage[section]{placeins}
\usepackage[T1]{fontenc}
\usepackage{caption}
\usepackage{subcaption}
\usepackage{enumerate}
\usepackage{color}
\usepackage[section]{placeins}
\usepackage{tikz}
\usetikzlibrary{calc,positioning,arrows}
\usepackage{listings}
\usepackage{longtable}
\usepackage{hyperref}  
\usepackage{gensymb}
\usepackage{cancel}
\usepackage{authblk}
\usepackage{array}
\usepackage{wasysym}
\usepackage{rotating}
\usepackage{xspace}
\usepackage{natbib}
\usepackage{hyphenat}
\usepackage{rotating}
\usepackage{pdflscape}
\usepackage{afterpage}
\usepackage{adjustbox}
\usepackage{verbatim}
\usepackage{multicol}

\newcommand{\OnlineDoc}{\href{https://macular.gitlabpages.inria.fr/macular/user_doc/Macular/main.html}{the online documentation page }}

\newcommand{\sSec}[2]{\subsection{#1 \label{Sec:#2}}}
\newcommand{\ssSec}[2]{\subsubsection{#1 \label{Sec:#2}}}

\newcommand{\pare}[1]{\left(\, #1 \, \right)}

\newcommand{\bra}[1]{\left[\, #1 \, \right]}

\newcommand{\Cell}[2]{{#1}_{#2}} 

\newcommand{\W}[4]{W^{\Cell{#1}{#2}}_{\Cell{#3}{#4}}}
\newcommand{\Synapse}[2]{{#1}_{#2}} 
\newcommand{\Inp}[2]{\vec{\cI}^{(\Cell{#1}{#2})}} 
\newcommand{\Inps}[2]{\cI^{(\Cell{#1}{#2})}_{syn}} 
\newcommand{\Vnps}[2]{V^{(\Cell{#1}{#2})}_{syn}} 
\newcommand{\FRnps}[2]{FR^{(\Cell{#1}{#2})}_{syn}} 
\newcommand{\vf}[2]{\vec{\cF}^{(\Cell{#1}{#2})}} 
\newcommand{\vX}[2]{\vec{\cX}^{(\Cell{#1}{#2})}} 
\newcommand{\vmu}[2]{\vec{\mu}^{(\Cell{#1}{#2})}} 
\newcommand{\Is}[4]{I^{(\Cell{#1}{#2} \to \Cell{#3}{#4})}_{syn}} 
\newcommand{\Vs}[4]{V^{(\Cell{#1}{#2} \to \Cell{#3}{#4})}_{syn}} 
\newcommand{\FRs}[4]{FR^{(\Cell{#1}{#2} \to \Cell{#3}{#4})}_{syn}} 

\newcommand{\toS}[1]{\stackrel{#1}{\to}}
\newcommand{\cF}{{\mathcal F}}
\newcommand{\cG}{{\mathcal G}}
\newcommand{\cK}{{\mathcal K}}
\newcommand{\cI}{{\mathcal I}}
\newcommand{\cN}{{\mathcal N}}

\newcommand{\cX}{{\mathcal X}}
\newcommand{\K}[2]{\cK_{\Cell{#1}{#2}}}  

\newcommand{\G}[3]{\cG^{({#1} \toS{#3} {#2})}} 

\newcommand{\no}{}

\newcommand{\yes}{\CIRCLE}

\definecolor{eclipseStrings}{RGB}{42,0.0,255}
\definecolor{eclipseKeywords}{RGB}{127,0,85}
\colorlet{numb}{magenta!60!black}
\lstdefinelanguage{json}{
    basicstyle=\normalfont\ttfamily,
    commentstyle=\color{eclipseStrings}, 
    stringstyle=\color{eclipseKeywords}, 
    numbers=left,
    numberstyle=\scriptsize,
    stepnumber=1,
    numbersep=8pt,
    showstringspaces=false,
    breaklines=true,
    frame=lines,
    string=[s]{"}{"},
    comment=[l]{:\ "},
    morecomment=[l]{:"},
    literate=
        *{0}{{{\color{numb}0}}}{1}
         {1}{{{\color{numb}1}}}{1}
         {2}{{{\color{numb}2}}}{1}
         {3}{{{\color{numb}3}}}{1}
         {4}{{{\color{numb}4}}}{1}
         {5}{{{\color{numb}5}}}{1}
         {6}{{{\color{numb}6}}}{1}
         {7}{{{\color{numb}7}}}{1}
         {8}{{{\color{numb}8}}}{1}
         {9}{{{\color{numb}9}}}{1}
}

\newcolumntype{L}[1]{>{\raggedright\let\newline\\\arraybackslash\hspace{0pt}}m{#1}}
\newcolumntype{C}[1]{>{\centering\let\newline\\\arraybackslash\hspace{0pt}}m{#1}}
\newcolumntype{R}[1]{>{\raggedleft\let\newline\\\arraybackslash\hspace{0pt}}m{#1}}
\newcommand{\hrefi}[2]{\href{#2}{#1}}

\graphicspath{{Figures/}{}}

\title{Macular: a multi-scale simulation platform for the retina and the primary visual system}
 \author[1]{Bruno Cessac}
 \author[2]{Erwan Demairy} 
\author[1]{Jérôme Emonet} 
\author[1,2]{Evgenia Kartsaki}
\author[2]{Thibaud Kloczko} 
\author[2]{C\^ome Le Breton}  
\author[2]{Nicolas Niclausse}
\author[1,4]{Selma Souihel} 
\author[2]{Jean-Luc Szpyrka}
\author[2]{Julien Wintz}
  \affil[1]{\small{Université Côte d’Azur, Inria\\
Biovision team and Neuromod Institute\\
Sophia Antipolis, France}}
\affil[2]{Service d'expérimentation et développement - SED-Inria\\
Sophia Antipolis, France}
\affil[3]{P16 - Programme IA\\ 
  Inria Rocquencourt\\ 
  Domaine de Voluceau\\ 
  78150 Le Chesnay-Rocquencourt, France\\
  }

\begin{document}

\maketitle

\begin{abstract}
We developed Macular, a simulation platform with a graphical interface, designed to produce \textit{in silico} experiment scenarios for the retina and the primary visual system. A scenario consists of generating a three-dimensional structure with interconnected layers, each layer corresponding to a type of “cell” in the retina or visual cortex. The cells can correspond to neurons or more complex structures (such as cortical columns). The inputs are arbitrary videos. 
The user can use the cells and synapses provided with the software, or create their own using a graphical interface where they enter the constituent equations in text format (e.g., LaTeX). They also create the three-dimensional structure via the graphical interface. Macular then \textit{automatically} generates and compiles the C++ code and generates the simulation interface. This allows the user to view the input video and the three-dimensional structure in layers. It also allows the user to select cells and synapses in each layer and view the activity of their state variables. Finally, the user can adjust the phenomenological parameters of the cells or synapses via the interface. We provide several example scenarios, corresponding to published articles, including an example of a retino-cortical model. Macular was designed for neurobiologists and modelers, specialists in the primary visual system, who want to test hypotheses \textit{in silico} without the need for programming. By design, this tool allows natural or altered conditions (pharmacology, pathology, development) to be simulated.
\end{abstract}

\section{Introduction}

Our visual system has an extraordinary capacity. It is capable of converting the flow of photons emanating from our environment into a flow of electrical impulses that can be interpreted by our brain and our consciousness. This allows us to react quickly and effectively to the movements and changes that are constantly occurring around us. The process starts in the retina. This organ owes its efficiency, on the one hand, to its layered structure, composed of different types of neural layers—from photoreceptors to ganglion cells—connected by specific synapses to form neural circuits that respond to local visual characteristics. On the other hand, the retina is a fundamentally dynamic object. Just as much as its structure, the variety of time scales involved in neural and synaptic processes are essential for  enabling the retina to encode visual information efficiently, taking into account the fact that our environment is constantly in motion.

Our knowledge of the retina is essentially based on experimentation. For over a century, this has enabled us to characterise its structure and the way in which a multitude of specific circuits  work together to generate a reliable representation of visual scenes. However, given such a high level of complexity, involving a wide range of time scales, experimentation alone cannot provide a holistic description. Furthermore, experiments are costly in terms of resources, time and energy. In this context, numerical simulation combined with modelling are valuable assets. Even though no simulation is currently capable of reproducing the behaviour of a complete retina, they can reproduce the behaviour of a particular circuit or combination of circuits, explore hypotheses and vary physiological parameters that are difficult to access experimentally. It is therefore natural that numerous retinal simulation platforms have been developed (see section \ref{Sec:Alternatives} for a non exhaustive list).

While the simulation platform we are presenting here, Macular, fits into this perspective, it nevertheless differs significantly from existing platforms. Furthermore, although it includes the VirtualRetina simulator developed by members or former members of our group \citep{wohrer-kornprobst-etal:09}, it differs from it in several ways. Macular was actually designed with several requirements in mind. First, it is intended for experimenters or modellers, with no programming knowledge, who would like to be able to simulate situations that interest them. Macular offers an interface that allows them to enter equations (e.g. in LaTeX) characterising the dynamics of specific neurons or synapses, and then organise these neurons/synapses into a multi-layered hierarchical structure mimicking the organisation of the retina. Without using a programming language, they can then generate a simulation of this structure. Furthermore, the parameters of these equations, corresponding for example to physiological parameters, can be modulated via an interface. This makes it possible to vary "manually" e.g. the conductance of an ion channel or the intensity of a synaptic connection. Thanks to this flexibility, Macular is not limited to modelling the retina alone, but also allows thalamic or cortical extensions to be added. An example of a cortical extension is provided in section \ref{Sec:Cortex}. Finally, with Macular, we wanted to be able to study the response to realistic visual stimuli, such as those used in experiments. Thus, Macular accepts films as "visual" input (this feature is inherited from Virtual Retina).
However, the retina does not always receive visual input. During development, before birth when the photoreceptors are inactive, there is nevertheless electrical activity (retinal waves) that we wanted to be able to simulate. Another situation concerns retinal prostheses, where the "input" is an electrical stimulation that is also possible with our simulator.

Macular runs on the three main operating systems: Linux, Mac and Windows.  The aim of this article is to provide a brief overview of this platform, bearing in mind that more comprehensive online documentation is available in \OnlineDoc.  
 The article is structured as follows. In section \ref{Sec:GeneralPresentation} we provide a general presentation of Macular, its spirit and structure. In section  \ref{Sec:GUI} we present the GUI and its main features. In section 
\ref{Sec:MTE} we  expose how to create cells or synapses of new type using the Macular Template Engine. Macular also has a batch version presented in section \ref{Sec:Batch}. 
Section \ref{Sec:Examples} provides then a few examples of use case including the simulation of retinal waves based on a model published in \citep{cessac-matzakou-karvouniari:22,karvouniari-gil-etal:19} and retino cortical model published in \citep{emonet-cessac:25,emonet-souihel-etal:25}. Section \ref{Sec:Alternatives} shortly presents existing simulators in the spirit of Macular and compares them to our platform. 

\section{General Presentation}\label{Sec:GeneralPresentation}

\sSec{Installation}{Installation}

Macular is a free software (GPL), written in C++, with the licence number\\
IDDN.FR.001.020016.001.S.P.2022.000.31235. It can be freely downloaded
at this  \href{https://macular.gitlabpages.inria.fr/macular/user_doc/Macular/main.html}{url} by following the instructions given at this page. A git repository is avalaible  \href{https://gitlab.inria.fr/macular/macular}{here}.

\sSec{Overview}{Overview}

The entire structure and conception of Macular relies on the following observation. The bio-physics of the retina and of the visual system can be modelled, with a very good accuracy, by
(partial or ordinary) differential equations. These equations are, in general, complex, non linear, with many degrees
of freedom, multiple space and time scales, and have non stationary (visual) inputs. Still, it is
possible to simulate them using adapted numerical schemes and structures.

\begin{figure}[!ht]
\centering
\includegraphics[width=0.45\textwidth,height=0.3\textheight]{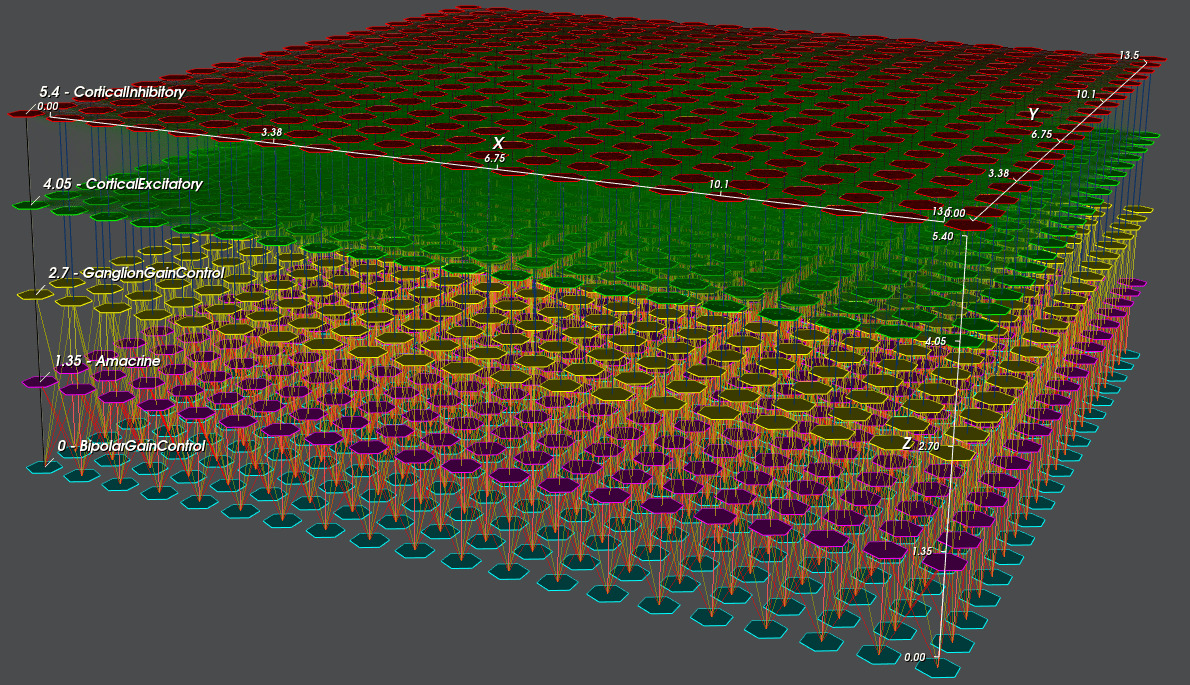}
\caption{\textbf{The multi-layer structure of Macular.} Here, we show a scenario involving three retinal layers and two cortical layers further described in section \ref{Sec:Cortex}. This scenario has been used in the papers \citep{emonet-cessac:25,emonet-souihel-etal:25}.}
\label{Fig:MacularGraph}
\end{figure}

Macular is organised into a layered structure that mimics the multi-layer organisation of the visual system (Fig. \ref{Fig:MacularGraph}). It is fed by visual inputs (movies) then processed by this multi-layer structure. 
At the heart of Macular are objects called "Cells", inspired by biological neurons, but more general. A "Cell" can also be a group of neurons of the same type, a neural field generated by a large number of neurons (for example a cortical column), or even an electrode in a retinal prosthesis. To differentiate
biological cells from Macular Cells we will use a capital in this latter case. More generally, Macular
objects like Synapses, Currents will be designed with a capital.
A Cell is defined by internal variables (evolving over time), internal parameters (adjusted by cursors), a dynamic evolution (described by a set of differential equations) and inputs. Inputs can come from an external visual scene or from other synaptically connected cells. Synapses are also Macular objects defined by specific variables, parameters, and equations. Cells of the same type are connected in layers according to a graph with a specific type of Synapses (intra-Layer connectivity). Cells of a different type can also be connected via Synapses (inter-Layer connectivity).

All the information concerning the types of Cells, their Inputs, their Synapses and the organization of the Layers are stored in a file of type .mac (for "macular") defining what we call a "scenario". Different types of scenarios are offered to the user, which they can load and play, while modifying the parameters and viewing the variables. More generally, Macular is built around a central idea: its use and its graphical interface can evolve according to the user's objectives, so, the user can design their own scenarios, i.e. define their own Cells, Synapses, Layers, using a specific template, the Macular Template Engine. This template, and more generally, Macular, has been designed so that the user does not need to use computer programming to run their simulations. 

Although Macular targets simulations of the retina, it  is not limited to it. It is designed to propose and test models of the visual system, where, for example, Cells represent cortical columns in a mean-field model. However, Macular is, by no way, intended to simulate the retina or the early visual system \textit{as a whole}. Instead, it is designed to check hypotheses on \textit{specific} aspects of the visual system, try and reproduce specific experiments \textit{in silico}. It is a tool for modellers and experimentalists. Especially, one can play the same stimuli as experimentalists and then record the response of Cells and Synapses of the model Layers. This is why the notion of scenario built by the user is central. From this perspective, note that generating a model or a scenario requires to have a clear idea of the equations to use, their parameters, and last but not the least, a coherent set of physical units. Thus, proposing a realistic scenario requires an important phase of design. 

\subsection{Units}\label{Sec:Units}

Macular uses a set of physical units listed in table \ref{Tab:Units}. There is a default system of units, shown in the second column of the Table.  Macular converts the units of the user's model to the default units for computations and then reconvert it in the user's units for plots.  Note that space scales have $3$ possible "modalities": distance, angle or pixels (see \OnlineDoc for more detail). We note however that Macular does not check that the user's units are coherent, in contrast e.g. to \href{https://briansimulator.org/}{BRIAN} \citep{goodman-brette:08}.  

\begin{table*}[!ht]
\begin{center}
    \begin{tabular}{ || p{4cm} | p{5cm} | p{5cm} || }       			   \hline
    \textbf{Physical quantity} 	& 	\textbf{Default Macular Units}  & \textbf{Other Possible Units}					\\ \hline
    Time ($\tau$)   & second (s) & milli-second (ms) \\ \hline
    Voltage ($V, E$)   & milli-volts (mV) & volts (V) \\ \hline
    Electric current ($I$)   & pico-ampère (pA) & nano-ampère (nA), micro ampère per $cm^2$ ($\mu/cm^2$)\\ \hline
    Electric conductance ($g$)   & nano-siemens (nS) & pico-siemens (pS), milli-siemens per $cm^2$ (mS/$cm^2$) \\ \hline
    Distance ($\sigma$)   & millimeters (mm), degrees ($\degree$), pixels (px) & micro-meters ($\mu m$) \\ \hline
    Capacitance ($C$)   & nano-farad (nF) & pico-farad (pF), micro-farad per $cm^2$ ($\mu F$/$cm^2$) \\ \hline
    Frequency ($\nu,f$)   & hertz (Hz) & kilo hertz ($kHz$) \\ \hline
    Molarity ($M$)   & nano-mol per liter (nM) & milli-mol per liter ($mM$), micro-mol per liter ($\mu M$) \\ \hline

    \end{tabular}
\end{center}
\caption{Physical units used in Macular. The first column displays the name of the main physical quantities used in models, with the letter that usually identifies them (in parentheses). In the second column, we show the default unit of these quantities in Macular. The third column presents the other units available in Macular. In the Macular interface (GUI) "micro" is denoted u instead of $\mu$. 
\label{Tab:Units}
}
\end{table*}

\sSec{Core Architecture}{CoreArchitecture}

We assume here that the reader knows the retina structure (for a very didactic introduction see e.g. \href{https://onlinebooks.library.upenn.edu/webbin/book/lookupid?key=olbp60030}{the web vision page} by Helga Kolb). 
 In the next lines, for simplicity with respect to the biological reality, we call \textit{OPL} (Outer Plexiform Layer) the retina region that contains photo-receptors (rods and cones) and Horizontal cells (HCs), and \textit{IPL} (Inner Plexiform Layer) the region which contains Bipolar cells (BCs), Amacrine cells (ACs) and Retinal Ganglion cells (RGCs). 
More generally, we extend the notion of Layers to models containing cortical populations, each population corresponding to a Layer. 


\subsubsection{Visual flow}\label{Sec:VR}


In Macular, the OPL is essentially represented by BCs receptive field (RF). Biologically, the RF of a BC is a region of the visual field (the physical space) in which stimulation alters its voltage (evokes a response of the Cell). This definition actually generalises to other retinal cells type like ACs or RGCs but we stick to BCs here. 
In Macular we model the receptive field of a BC $i$ of type $T$, $\Cell{T}{i}$, as a spatio-temporal kernel $\K{T}{i}(x,y,t)$, i.e. a function of space and time with a specific structure. This receptive field features the lateral inhibition coming from horizontal cells in the form of a difference of Gaussians.

The linear response of the RF to a visual stimulus is then given by a space-time convolution (see e.g  \href{https://en.wikipedia.org/wiki/Receptive_field}{this web page}).

In Macular, convolutions are computed using a fast method called Deriche filters \citep{deriche:87} and are handled by the Virtual Retina simulator, developed by Wohrer and Kornprobst \citep{wohrer-kornprobst-etal:09}, integrated in Macular.
As a consequence, filters have a \textit{spherical symmetry}. This limitation is further discussed in the conclusion section. Stimuli are considered as levels of gray between $[1,255]$. We do not handle color in Macular. A detailed description of the Virtual Retina implementation in Macular can be found \href{https://macular.gitlabpages.inria.fr/macular/user_doc/Macular/main.html#appendix}{here}.

\subsubsection{Cells}\label{Sec:Cells}

We now define more specifically what  Macular Cells are. A Cell is denoted $\Cell{T}{i}$ where $T$ is called the "Cell type" and $i$ is the index labelling Cells of type $T$. A Cell type can refer either to the cell's biological terminology (e.g. bipolar or amacrine retina cell layers), to subtypes within these general cell layers (e.g. starburst amacrine cell), or to its functionality (e.g ON cells). But, as already mentioned, a Macular Cell does not necessarily correspond to a biological cell. It can be, for example, a region in the cortical space (e.g. a cortical column) corresponding to a mean-field average over thousands of neurons (see section \ref{Sec:Cortex}). A glossary of (default) Cell types existing in Macular is given in table \ref{Tab:Cells}.

The Cell $\Cell{T}{i}$ is identified by:

\begin{itemize}
    \item \textbf{An Input, $\Inp{T}{i}(t)$}. The Cell receives an entry which can be:
    \begin{enumerate}
        \item An external Input $\Inp{T}{i}_{ext}$. 
For example, an entry corresponding to the input from OPL (visual flow, i.e. the convolution of a movie with the OPL receptive field) to bipolar cell  $\Inp{T}{i}_{OPL}$ (defined in section \ref{Sec:VR}), or the electric current provided by an electrode $\Inp{T}{i}_{stim}$ (defined in section \ref{Sec:Electrodes}). 
   
        \item A synaptic input $\Inps{T}{i}(t)$ 
        corresponding to Synaptic connections with other Cells and defined in section \ref{Sec:Synapses}. In general, this contribution sums up the connection with several pre-synaptic cells.
    \end{enumerate}{}
    The input $\Inp{T}{i}(t)$ is in general the sum of several  contributions (e.g. OPL current and synaptic input).
    
    \item \textbf{A State}. This is an array $\vX{T}{i}$
    of \textit{variables} evolving in time and characterizing the Cell’s dynamical evolution. For example, State variables can be a membrane potential, activity - probability that an ion channel of a given type is open, concentration of neurotransmitter of a given type released by the cell, etc. 
    
    \item \textbf{A set of Parameters}. These are quantities that do not evolve in time but are nevertheless necessary to constrain the Cell evolution. They can, for example, correspond to conductances, reversal potentials, membrane capacitance, etc. They can be modified by the user, with sliders or by typing the value in a field. We denote by $\vmu{T}{i}$ the array of these parameters.
    
    \item \textbf{A function,} called Vector Field $\vf{T}{i}$, controlling the time evolution of Cells. Mathematically, $\vf{T}{i}$ is the vector field of the differential equation:
\begin{equation} \label{eq:VectorField}
    \frac{d \vX{T}{i}}{dt}=\vf{T}{i}(\vX{T}{i},\vmu{T}{i},\vec{I^{(i)}}(t)),
\end{equation}{}
and $\vf{T}{i}$ has the same dimension as $\vX{T}{i}$, the State vector.

\end{itemize}{}

\subsubsection{Pre-defined cell types}\label{Sec:Prefefined_cells}

There is a set of pre-defined Cells defined in Macular listed in Table \ref{Tab:Cells}. 
The user can create new Cells using the MacularTemplateEngine presented in section \ref{Sec:MTE}. 
Most of the predefined Cells in Macular (except the so called "CorticalCells" which actually physically correspond to cortical columns) are based on the generic equation for voltage:
\begin{equation} \label{eq:Generic_V_Isyn}
 C \frac{dV}{dt} \,=\, -\, g_L \, \pare{V \,-\, E_L} \,-\, \sum_X g_X \, \pare{V \,-\, E_X} \,+\, I_{syn} \,+\, I_{ext}
\end{equation}
where $g_L$ and $E_L$ respectively refer to leak conductance and leak reversal potential, $g_X$ and $E_X$ correspond to ionic current contributions, $I_{syn}$ is the synaptic current discussed in section \ref{Sec:Synapses} and $I_{ext}$ is the input Current. 
Another form, also used in Macular, is:
\begin{equation} \label{eq:Generic_V_Vsyn}
\frac{dV}{dt} \,=\, -\, \frac{V \,-\, E_L}{\tau_L} \, -\, \sum_X \, \frac{V \,-\, E_X}{\tau_X} \,+\, V_{syn} \,+\, V_{ext}.
\end{equation}
It corresponds to \eqref{eq:Generic_V_Isyn} setting $\tau_L=\frac{C}{g_L}$, $\tau_X=\frac{C}{g_X}$. 
The term $V_{syn}$ appearing in \eqref{eq:Generic_V_Vsyn} corresponds to the synaptic input, explained in section \ref{Sec:Synapses}. Note that it does not have the dimension of a voltage. Its dimension is $mV \, s^{-1}$ and would correspond, from \eqref{eq:Generic_V_Isyn}, to $\frac{I_{syn}}{C}$. We adopted the letter $V$ for simplicity. 
The same remark holds for the physical dimension of $V_{ext}$.  $I_{ext}$ and $V_{ext}$ correspond to different stages of integration in the OPL.

\begin{table*}
\centerline{{\tiny

\begin{tabular}{ || p{5.5cm} | p{8cm} | p{6cm} ||}       			   \hline
Cell name in Macular	& 	Equation & Comment				\\ \hline
{\scriptsize    macularCellAmacrine, macularCellBipolar}		& 	{\scriptsize    
$
\begin{array}{lll}
&\\
&\frac{d V}{d t} = -\frac{V-E_L}{\tau} \, + \, V_{syn}
\end{array}
$ } & {\scriptsize    Linear cell with synaptic input ($V_{syn}$) and characteristic time $\tau$. $E_L$ is the leak reversal potential.} 
\\ \hline
{\scriptsize    macularCellAmacrineLinearPharma, macularCellBipolarLinearPharma, macularCellGanglionLinearPharma}		& 	
{\scriptsize   
$
\begin{array}{lll}
&\\
&\frac{d V}{d t} = -\frac{g_L+g_P}{C}V +  \frac{I_{syn}}{C} \,+\, \frac{g_L\, E_L\,+\,g_P\,E_P}{C}
\end{array}
$
}
& {\scriptsize  Linear cell with membrane capacitance $C$ and a tunable ion contribution e.g. corresponding to an injected drug where $g_P$:  conductance; $E_P$: Nernst potential, of the ionic channels sensitive to that drug (see \cite{kartsaki-hilgen-etal:24})} \\ \hline
{\scriptsize macularCellAmacrineGABA, macularCellAmacrineAMPA}		
&{\scriptsize   
$
\begin{array}{lll}
&\\
&\frac{d T}{d t} = -k_d \, T + \frac{k_p}{1+e^{(-(V-E_N)/\kappa_N)}}\\
&\frac{d V}{d t} = - \frac{V}{\tau_A} + \frac{I_{syn}}{C_A}  \\
&\frac{dn}{dt} = -\beta_{n} n +  \alpha_{n} \, T \, (1-n)
\end{array}
$
}
 &{\scriptsize  Linear Amacrine cell producing GABA (resp. AMPA)  with a quasi static production of neurotransmitter T and an activation variable n. From \cite{destexhe-mainen-etal:98}}
\\ \hline
{\scriptsize macularCellBipolarGainControl}  & 
{\scriptsize 
$
\begin{array}{lll}
    &\\
    &\frac{dA_B}{dt} \,=\, -A_B/\tau_{A_B} \,+\, h_B\, N_B(V)\\
    &\frac{dV}{dt} \,=\, -(V-E_L)/\tau_B \,+\, Vext/\tau_{ext} \,+\, V_{syn}
\end{array}
$
} &  {\scriptsize Bipolar Cell with a gain control controlled by a non linear function $N_B$ of the voltage V and of an activity variable $A_B$ (from \cite{berry-brivanlou-etal:99,chen-marre-etal:13,souihel-cessac:19}).} 
\\ \hline
{\scriptsize macularCellCorticalExcitatory, macularCellCorticalInhibitory}  & {\scriptsize The equations are too long to be written in this table. For further detail see \cite{zerlaut-chemla-etal:18,emonet-souihel-etal:24}} &  {\scriptsize  Respectively excitatory and inhibitory populations of a cortical column. Excitatory populations come from regular spiking cells (RS) and inhibitory populations from fast spiking cells (FS).}
\\ \hline
{
\scriptsize macularCellElectrode
}  
& {
\scriptsize 
$
\begin{array}{lll}
    &\\
&\frac{d V}{d t} = -\frac{V}{\tau} \, + \, \frac{I_{ext}}{C}
\end{array}
$ 
}
&  
{
\scriptsize Passive (low pass) electrode receiving an input, $I_{ext}$, corresponding to a local pixel average (see section \ref{Sec:Electrodes}).
} 
\\ \hline
{\scriptsize macularCellGanglionGainControl}	& 	{\tiny 
$
\begin{array}{lll}
&\\
&\frac{d V}{d t}= -\frac{1}{\tau_L}\,(V-V_L) + V_{syn} \,-\, \frac{g_{T}}{C_G}\,(V-V_T);\\
&\frac{d A_G}{d t}=	- \frac{A_G}{\tau_G} \, + \, H_G \, \cN_G(V)
\end{array}
$
}
&
{
\scriptsize Ganglion cells with gain control (from \cite{berry-brivanlou-etal:99,chen-marre-etal:13}) and  a tunable ion  contribution e.g. corresponding  to an injected drug. Firing rate is controlled by a non linear function $N_G$ of the voltage V and of an activity variable $A_G$.
}  	
\\ \hline
{
\scriptsize macularCellHodgkinHuxleyCurrent, macularCellHodgkinHuxleyVoltage
}
& {
\scriptsize 
From \cite{hodgkin-huxley:52}.
} &  
\scriptsize{
Hodgkin-Huxley neuron with the classical form \eqref{eq:Generic_V_Isyn} or with a voltage form \eqref{eq:Generic_V_Vsyn}.
} 
\\ \hline
{
\scriptsize macularCellMorrisLecar
}  
& {
\scriptsize 
From \cite{morris-lecar:81}.
} &  
\scriptsize{
Morris-Lecar neuron.
} 
\\ \hline
{
\scriptsize macularCellMorrisLecarAch
}  
& {
\scriptsize 
Used to feature Starburst Amacrine Cells. Parameters have been tuned according to the paper \cite{cessac-matzakou-karvouniari:22}
} &  
\scriptsize{
Morris-Lecar neuron producing Acetylcholine.
} 
\\ \hline
{\scriptsize macularCellSAC}  & {\scriptsize Used to feature Starburst Amacrine Cells during development. Parameters have been tuned according to the paper \cite{cessac-matzakou-karvouniari:22}} &  {\scriptsize Morris-Lecar neuron producing Acetylcholine with a potassium slow After Hyperpolarization current.} \\ \hline
\end{tabular}
}
}
\caption{Cell types pre-defined in Macular, listed in alphabetic order. New Cell types can be created using the Macular Template Engine (section \ref{Sec:MTE}). Once a variable has been introduced in the table, we do not repeat its definition. More detail can be found by clicking on the variable name in the Macular GUI}
\label{Tab:Cells}

\end{table*}

We distinguish $3$ main Cells subtypes, based on the mathematical implementation of the conductances $g_X$:
\begin{itemize}
    \item \textbf{Linear cells.} The conductances $g_X$ are constant, i.e. they do not depend on any variable. 
    \item \textbf{Rectified cells.} The conductances depend on voltage only, and take the form\\
    $g_X(V)\,=\, \lambda \, \cN_X(V)$ where $\lambda$ is a constant and:
    \begin{equation}\label{eq:Rectif}
    \cN_X(V) = \left\{\begin{array}{lll}
    V-\theta_X, \quad &\mbox{if} \,\, V > \theta_X;\\
    0, \quad &\mbox{otherwise}  
    \end{array} \right.,
    \end{equation}
    is a piecewise linear rectifier, $\theta_X$ being a voltage threshold.
    \item \textbf{Non-Linear cells.} The conductances depends non linearly on voltage and on potential additional variables like activation or inactivation variables. This is the case e.g. for cells inspired from Morris-Lecar \cite{morris-lecar:81} or Hodgkin-Huxley 
    model \cite{hodgkin-huxley:52}.
\end{itemize}

In addition, some Cell types have activation variables used for synaptic computation (see section \ref{Sec:Synapses}).
Note that there is no constraint for the user to stick to Cells of the form \eqref{eq:Generic_V_Isyn} or \eqref{eq:Generic_V_Vsyn}. They are free to develop their own using the Macular Template Engine (section  \ref{Sec:MTE}).

\subsubsection{Cell Layers}\label{Sec:Layers}
Macular is organised in Layers.  A Cell layer is a set of Cells of the same type $T$, where "same type" means that the Inputs, State vector, Parameters vector and Vector Field have the same mathematical expression. In this respect, Macular Layers are different from biological "layers" which can contain different cell types. Note that \textit{Cells in the same Layer share the same set of parameters}. The State values can differ, depending on the initial conditions and on the Input. 
In Macular, Cells are considered as points i.e. soma, axons, synapses of neurons are located at the same point. They are identified by an index (ID). Cells within a given Layer are organized in a two dimensional grid, and different Cell Layers are located on a $3$-dimensional space
with coordinates $(x,y,z)$. All Cells of type $T$ have the same $z$ coordinate. Thus, the Cell $\Cell{T}{i}$ has coordinates $(x_i,y_i,z_T)$ where the vertical coordinate $z_T$ parametrizes the Cell's type and the coordinates $(x_i,y_i)$  the position of Cell $i$ in the Layer $T$. All Layers have a common frame, with parallel axes in the $x,y$ directions and a common origin. 
Layers are represented as rectangles. The number of Cells in the horizontal and vertical direction might not be the same and each Layer may contain a different number of Cells. 

\subsubsection{Synapses}\label{Sec:Synapses}

Biological cells can be connected either by chemical synapses or by electric synapses (gap junctions). The synaptic contact between two cells involves complex dynamical processes such as calcium influx, release, diffusion and capture of neurotransmitter, opening or closing of ion channels resulting in electric currents modifying the membrane voltage of the post synaptic neuron. As well as in neuron modelling, these mechanisms are modelled by equations capturing different aspects of synapse dynamics.
In Macular, the objects called Synapses implements these aspects. 
A Synapse is noted $\Synapse{S}{k}$ where $S$ is called the "Synapse type" and $k$ is the index labelling Synapses of type $S$. The "type" of the Synapse refers here to a model, a set of equations, corresponding to a biological synapse, for example, a cholinergic synapse between two amacrine cells. The Synapse connects a pre-synaptic Cell $\Cell{T}{i}$ to a post-synaptic Cell $\Cell{T'}{j}$. \\

A Synapse is identified by:

\begin{itemize}
    \item \textbf{A set of Parameters.} These are quantities that do not evolve in time but constrain  the connectivity function of the Synapse. These can be conductance, connectivity weights, reversal potentials ... They can be modified by the user by sliders or by typing the value in a field. 
    \item \textbf{A function}, the mathematical representation of the synaptic connection. It could compute either a synaptic current ($\Is{T}{i}{T'}{j}$), a voltage ($\Vs{T}{i}{T'}{j})$ (i.e. a Post Synaptic Potential) or a firing rate ($\FRs{T}{i}{T'}{j})$. These quantities depend in general on the State vector of pre- and post-synaptic Cells. 
\end{itemize}

The predefined types of Macular synapses are listed in table \ref{Tab:Synapses}.

\begin{table*}[!ht]
\begin{tabular} { | p{6cm} | p{5cm} | p{6cm} |}       			   \hline
    Synapse name 	& 	Definition 	& Comment				\\ \hline
{\tiny    macularSynapseAcetylcholine}		& 	
%
{\tiny   
$
\begin{array}{lll}
     &  \\
     &\Is{T}{pre}{T'}{post}(t) \\
     &=\, - g_A \frac{A_{pre}^2}{\gamma_A + A_{pre}^2}.(V_{post} - V_A)
\end{array}$
}& {\tiny Model of Ach conductance for nicotinic receptors (from \cite{karvouniari-gil-etal:19}). $A_{pre}$, is the Ach  concentration emitted by the pre-synaptic cell (so the Cell type must contain this variable, for example a  macularCellSAC defined in table \ref{Tab:Cells}); $V_{post}$, voltage of the post-synaptic cell; $g_A$, Max Ach conductance; $\gamma_A$, half-activation constant; $V_A$, reversal potential for Ach.} \\ \hline
{\tiny   macularSynapseAmacrineToBipolar, macularSynapseAmacrineToGanglion, macularSynapseBipolarToAmacrine, macularSynapseBipolarToGanglion, macularSynapseLinearRectified} & 
%
{\tiny   
$
\begin{array}{lll}
     &  \\
     &\Vs{T}{pre}{T'}{post}(t)\\
     &= \,w^{pre}_{post} \, \cN_{pre}(V_{pre}-\theta_{pre})
\end{array}
     $
     }
     & \tiny{Rectified synapse (in mV/s). $w^{pre}_{post}$, synaptic weight from pre-synaptic to post-synaptic Cells; $\cN$, linear rectifier; $\theta_{pre}$, rectifying threshold (mV).}
 \\ \hline
{\tiny   macularSynapseAmacrineToBipolar, macularSynapseBipolarToAmacrine} & 
{\tiny   
$\Vs{T}{pre}{T'}{post}(t)\,=\,w^{pre}_{post} \,V_{pre}$
}
& \tiny{Linear synapse (in mV/s). } 
\\ \hline
{\tiny   macularSynapseBipolarGainControlToAmacrine, macularSynapseBipolarToAmacrine, macularSynapseBipolarPooling} & 
{\tiny   
$\Vs{T}{pre}{T'}{post}(t)\,=\,w^{pre}_{post} \, preBipolarResponse$
}& \tiny{The post synaptic voltage is proportional to the pre-synaptic voltage via a synaptic weight $w^{pre}_{post}$. The term "preBipolarResponse" depends on the Macular Cell type.}
\\ \hline
{\tiny   macularSynapseGapJunctionVoltage} & {\tiny   
$\Vs{T}{pre}{T'}{post}(t)\,=$

$-w_{gap}.(V_{post} - V_{pre})$} & {\tiny Passive gap junctions where $w_{gap}$ is expressed in $nS/nF = Hz$.} \\  \hline
{\tiny    macularSynapseCorticalExc\_to\_CorticalExc, macularSynapseCorticalExc\_to\_CorticalInh, macularSynapseCorticalInh\_to\_CorticalExc, macularSynapseCorticalInh\_to\_CorticalInh} & 
{
\tiny $\nu_{post} \,= \, w^{pre}_{post}. \nu_{pre}$
}&  {\tiny From \cite{zerlaut-chemla-etal:18}, where $\nu_{pre}, \nu_{post}$ are the firing rates of the pre/post-synaptic Cell (corresponding here to a cortical column) and $w^{pre}_{post}$ the gaussian weigth between pre/post-synaptic Cell.
}
\\ \hline
{\tiny  macularSynapseGABA\_A, 
macularSynapseAMPA  } & 
{\tiny
$
\Vs{T}{pre}{T'}{post}(t)\,=\,-\,g \,n\, (V_{pre}-E)
$
}
& {\tiny Here $n$ is an activation variable as produced by the Cells macularCellAmacrineGABA,  macu-larCellAmacrineAMPA. 
} 
\\ \hline
{\tiny  macularSynapseRetinoCortical } & 
{\tiny
$
FR_{syn}^{T_{pre} \to T'_{post}}(t) = weight . \frac{density_{retine}}{density_{cortex}} . preFiringRate
$
}
& {\tiny Synapse connecting the retina to the cortex \cite{souihel:19} where preFiringRate is the output firing rate of ganglion cells. It is multiplied by a factor corresponding to the ratio between the retinal density and the cortical density. Respectively $400 \, mm^{-2}$ and $4000 \, mm^{-2}$ in the model \cite{emonet-souihel-etal:24}. }  \\ \hline
\end{tabular}
\caption{Synapses type pre-defined in Macular. New Synapses type can be created using the Macular Template Engine (section \ref{Sec:MTE}).} 
\label{Tab:Synapses}

\end{table*}

In the Macular Graph Generator (see section \ref{Sec:Graph_Generator}), the user specifies the Cell's type in each Layer and select the Synapses' type inside a Layer (intra-Layer Synapses) or between Layers (inter-Layers Synapses). There can be several types of intra- or inter-Layers Synapses in the simulation.


A post-synaptic Cell receives in general many Inputs from different Cells of different  types. So, the general form of the Synaptic Current $\Inps{T}{i}(t)$ introduced in section \ref{Sec:Cells}, eq. \eqref{eq:Generic_V_Isyn}, is:
\begin{equation}\label{eq:Synaptic_Current}
I_{syn} \equiv \Inps{T}{i}(t)= \sum_{T'} \, \sum_{j \in T'} \Is{T}{i}{T'}{j}(t),
\end{equation}
where the first summation holds on the Cells $j$ of type $T$ pre-synaptic to Cell $i$  and the second summation holds on Cells Layers. 
The same formulation holds in the case of the voltage representation, introduced in section \ref{Sec:Cells}, eq. \eqref{eq:Generic_V_Vsyn}:
\begin{equation}\label{eq:Synaptic_Voltage}
V_{syn} \equiv \Vnps{T}{i}(t) = \sum_{T'} \, \sum_{j \in T'} \Vs{T}{i}{T'}{j}(t),
\end{equation}
or a firing rate input used e.g. for retino-cortical Synapses:
\begin{equation}\label{eq:Synaptic_Firing_RTte}
\FRnps{T}{i}(t) = \sum_{T'} \, \sum_{j \in T'} \FRs{T}{i}{T'}{j}(t).
\end{equation}

By default, synapses in Macular are instantaneous, i.e. there is no delay between the emission of a signal at the pre-synaptic neuron and its arrival at the post-synaptic one. It is nevertheless possible to add a delay to a Synapse type. For this, the user has to create a speed parameter called "conduction\_velocity" in the Synapse type. Macular compute the synaptic delay based on the equation :
\begin{equation}\label{eq:delay}
delay_{syn} \,=\, \frac{d_{syn}}{v_{C}},
\end{equation}
where $d_{syn}$ is the distance between the two neurons (e.g. the length of the axons), and $v_C$ the conduction velocity. 


\subsubsection{Graph}\label{Sec:Graph}
Synapses define a natural notion of intra- and inter-layer connectivity. If the Cell $\Cell{T}{i}$ is pre-synaptic to Cell $\Cell{T'}{j}$, with a Synapse of type $S$, we note $\Cell{T}{i} \toS{S} \Cell{T'}{j}$ the  oriented edge featuring this connection.  The set of edges of type $S$ defines a directed graph $\G{T}{T'}{S}$. This graph features the set of synaptic connections of type $S$, from Layer $T$ to Layer $T'$. If $T=T'$ we speak of "intra-Layer connectivity" of type $S$, and "inter-Layer" if $T \neq T'$. 
 Between two Layers there may exist several type of synaptic connections and a Cell can be a source or a target to different types of Synapses (e.g. an AC can connect a BC through a glycinergic synapse and a gap junction). 
 
 In this frame, Cell $i$ has coordinates $x_i,y_i$ in its Layer, while Cell $j$ has coordinates $x_j,y_j$ in its Layer. The distance between these two cells is $d(i,j)=\sqrt{\pare{x_i-x_j}^2+\pare{y_i-y_j^2}}$,  the two dimensional Euclidean distance. That is, we do not consider the vertical distance between different Layers. 
Two Cells are nearest neighbours if their distance is the smallest strictly positive distance.  

In Macular, there are 6 types of connectivity that a graph can implement between two Layers:
\begin{itemize}
\item \textbf{One-to-one (Inter-Layers).} A Cell is connected to the Cell at zero distance in another Layer.
This type of connectivity requires that these Layers have the same number of Cells.
\item \textbf{Nearest neighbours  (Inter- and Intra-Layers).} A Cell is connected to its 4 nearest neighbours.
\item \textbf{Neighbour 4 + 1 (Inter-Layers).} A Cell is connected to 4 nearest neighbours and to the Cell at distance zero.
\item \textbf{Radius neighbours  (Inter- and Intra-Layers).} A Cell is connected to neighbours Cells within a certain radius (excluding Cell at distance zero). The synaptic weights are constant within this radius.
\item \textbf{Gaussian (Inter- and Intra-Layers).} The synaptic weight between the pre-synaptic and the post-synaptic Cell depends on their distance $d(i,j)$, via a Gaussian profile:
\begin{equation}\label{eq:GaussianPooling}
   \W{pre}{}{post}{} \,=\, \frac{e^{-\frac{d(i,j)^2}{2 \,\sigma_{p}^2}}}{2 \pi \,\sigma_{p}^2}
\end{equation}
%
In this case, we have connectivity to the Cell at distance $0$. 
\item \textbf{Fully Connected (Inter-Layers).} A Cell is connected to all Cells with constant synaptic weights.
\end{itemize}

Among all of these connectivity types, Gaussian and Nearest Neighbours are currently the only ones used to connect Cells from the same Layer.

%


\subsubsection{ODE solver}\label{Sec:ODESolver}

Macular integrates ordinary differential equations (ODE) using the General Scientific Library (GSL)  (See the  \href{https://www.gnu.org/software/gsl/doc/html/ode-initval.html}{GSL online documentation}). The library provides a variety of low-level methods, such as Runge-Kutta and Bulirsch-Stoer routines, and higher-level components for adaptive step-size control. By default the method used in Macular is Runge Kutta of order 4 (RK4).
Note therefore that the current implementation of Macular is not adapted to simulate evolution with noise (which would require specific stochastic integrators). 
The Macular GUI menu allows the selection of different integration methods: RK2, RK4, RK45, RK8, RKCK, RK1imp, RK2imp, RK4imp, BSIMP, ADAMS, BDF. See the  \href{https://www.gnu.org/software/gsl/doc/html/ode-initval.html?highlight=rk4#}{online documentation of the GSL}
for detail on these methods.

\subsubsection{Electrodes stimulation.}  \label{Sec:Electrodes} Retinal implants are electronic devices surgically attached to the
retina. They substitute to defective cells in order to partially restore vision. Images acquired by a "camera + processor" system are encoded and sent as pulses to a matrix of electrodes. It then stimulates the still functional cells of the retina in order to reproduce a luminous impression. 

We have implemented in Macular a simplified version of this process. Electrodes are considered as "Cells" (type macularCellElectrode). They are quite simplified with respect to real electrodes as they are just low pass filters, but the user can extend their definition using more complex equations and the MacularTemplateEngine facilities (section \ref{Sec:MTE}). A retinal prosthesis is a matrix of electrodes which becomes, in Macular, a matrix of "macularCellElectrode".

The "camera + processor" processing is featured by averaging the pixels around the location of a given macularCellElectrode in a region whose size is the image size in pixels divided by the number of electrodes. This averaging provides the macularCellElectrode input. This functionality is obtained by selecting "Prosthesis" in the "WorkerSetting" (see section \ref{Sec:Simulator}). 

\section{The Macular GUI}\label{Sec:GUI}

Macular has a Graphical User Interface (GUI) with a large panel of possibilities such as the visualisation of the Cells Layers in 2D or 3D, the monitoring of specific Cells' State variable $\dots$. 
The majority of the elements in the Macular GUI have a small embedded documentation appearing when  pointing the mouse on it.


\sSec{Views}{Views}

When opening Macular a panel appear showing up 4 buttons corresponding to different views.

\begin{itemize}
    \item \textbf{3D view} creates "canvas" object that provide a layered, customizable, view of the simulation. 
    
    \item \textbf{Layered view} provides a set of 2D views "Views2D", one for each Layer, and is customizable.
    
     \item \textbf{Plot views} allows to generate a Plot2D object to monitor the time evolution of specific Cells variables.
     
       \item \textbf{Stimulus.} When an image or a video is played this option allows to see the stimulus. 
\end{itemize}

Several views can be simultaneously open.

\sSec{The Simulator}{Simulator}

\ssSec{The Configuration panel}{ConfigPanel}
On the left of the GUI 
a list of icons are visible. 
This is the configuration panel, respectively corresponding to the following functions. 

%
%
%
\begin{itemize}
 
    \item \textbf{Selection}. 
    The user can select which output they want to record in their simulation. 
    
    \item \textbf{Video Input}. 
   The command \textbf{Browse Stimulus} loads a visual stimulus in the form of a movie in the formats .mp4, .mkv, .avi. This stimulus will be played when running the simulation. 
 
   \item \textbf{Graph Input}. The command  \textbf{Browse Graph} loads a .mac (mac, for "Macular") file containing a Macular graph (see section \ref{Sec:Graph_Generator} for a description of the .mac files).

    \item \textbf{Worker settings}. 
    \textbf{Input} selects a Worker, namely a setting of functionalities to run the simulation according to the type of visual input. 
    The options are:
    \begin{enumerate}[(1)]
        \item "Visual Flow". The menu essentially contains Parameters shaping the Receptive Field filter (see section \ref{Sec:VR}). 
		\item "Prosthesis". Here one parametrizes the setting for retinal prostheses.
		\item "None". Here, there is no input. 
     \end{enumerate}     
	\item
	\textbf{Simulation parameters.} This menu allows to further parametrize the simulation.
%
	\item
	\textbf{Controls} allows to run, save, and reset the simulation.
	\item
	\textbf{Parameters of Cells and Synapses.} Here, one can select a predefined Cell or Synapse type.

%
	\item
	\textbf{Configuration} allows the user to select the appearance of the GUI (colors of the background, fonts) and to toggle advanced parameters selected in the parameters visibility view.

\end{itemize}


\ssSec{The graph Generator}{Graph_Generator}

The Graph Generator allows the user to create layers of Cells, with a given connectivity using the existing Cell/Synapse types (for the creation of new Cell types see section \ref{Sec:MTE}). An example of Graph creation is provided \href{https://macular.gitlabpages.inria.fr/macular/user_doc/Macular/main.html#examples}{here}.  As exposed in section \ref{Sec:Graph} a Graph is a mathematical structure made up of vertices (Cells), connected by intra- and inter-layer edges (Synapses). In Macular, a graph is implemented as a C++ object. The data necessary to run the simulation are saved in two files. The first one, with  the extension .mac  (mac, for Macular) contains the number of Cells, the number of Synapses the type of each Cell with its coordinates, the type of each Synapse and the Cells it connects to, and, finally the initial value of each variable. The second one, with an extension .json, contains the information about model parameters.
More details can be found in \OnlineDoc.

\section{The Macular Template Engine}\label{Sec:MTE}

The  Macular Template Engine (MTE) allows the user to manage existing Cells and Synapses types, to create new Cells and Synapses types, or to suppress them. Then, MTE generates automatically (i.e. without need of writing code) a set of C++ files.  After any change in MTE, the user has to press the "Write C++ files" and re-compile Macular using the "Build" button. \\

 \textbf{Important notice.}  Modifying existing Cells or Synapses will replace the user existing files, except for protected Cell and Synapse types. Indeed, some Macular Cell or Synapse type are protected: they contain a ".lock" extension at the end of its json file name. They can not be modified by the MTE. Currently, only two macular Cells are protected (macularCellCorticalExcitatory and macularCellCorticalInhibitory). \\

In more detail, the main features of the MTE are: 

\begin{itemize}
    \item \textbf{Loading} the Cells and Synapses types already existing in the directory share/macular/app/macularTemplateEngine/json/ (on Windows, the share directory is a subdirectory of the \emph{Library} directory). 
    
    \item \textbf{Creating and editing new Types.} This allows the user to create new Cells/Synapses types or to edit unprotected Cell/Synapses types with specific parameters, auxiliary functions and vector field equations.
    
    \item \textbf{Deleting existing Types.} This allows the user to suppress any unprotected Cell and Synapse type. The corresponding .json file (inside share/macular/app/macularTemplateEngine/json/ subfolder ) is suppressed.
    
    \item \textbf{Write C++ Files}: With this functionality, once the new Cells/Synapses types are saved in .json Files, the MTE will generate new C++/CMake files with the parameters, functions and equations specified in the .json files. For this purpose, Macular uses Python scripts  to generate the files based on C++ templates. This script fills in the required data in the C++ templates from the .json files, for each Cell and Synapse. The user can finally re-compile Macular  in order to add these new Cells/Synapses types to the Simulator and the Graph Generator. Note that this operation is done by simply pressing the button "Build". The source files for the generated Cells and Synapses are located in /macular/share/macular/src/macularCore/, assuming that macular is installed in /macular. 
\end{itemize}{}{}

The MTE is run by typing the shell command \texttt{bin/macularTemplateEngine  \&} in the main directory (Linux) (on MacOS, one runs the file macular.app in bin, and on Windows the binary should be available on your desktop after installation). An example of usage can be found \href{https://macular.gitlabpages.inria.fr/macular/user_doc/Macular/main.html#examples}{here}.

\section{Macular Batch}\label{Sec:Batch}
All the features available in the Macular GUI can be used in the batch version of Macular. This version is run from the main directory by typing the command "./bin/macular-batch -f path\_session\_file.json" in a shell/terminal (on Windows, it's Library/bin/macular-batch). The -f or --file argument is the only one mandatory. It requires the path to a Macular session json file. There are other options described in the \OnlineDoc.


\section{Examples}\label{Sec:Examples}

Here, we provide a few examples of simulations with scenarios included in the Macular release. 
For more detail see \OnlineDoc.

\sSec{Retinal Waves}{RetinalWaves}

This scenario, available in the directory\\
\emph{macular/examples/Scenario1\_RetinalWaves}  (on windows, it is located in \emph{examples}, at the top of the macular installation directory) provides a simulation of retinal waves occurring during the development of the visual system. It involves Starburst Amacrine Cells (SAC) which are sporadically synchronizing producing waves of bursting activity (see \citep{karvouniari-gil-etal:19,cessac-matzakou-karvouniari:22} for more detail on the model and references therein about developmental retinal waves).   

\begin{figure*}
\begin{center}
\includegraphics[width=\textwidth]{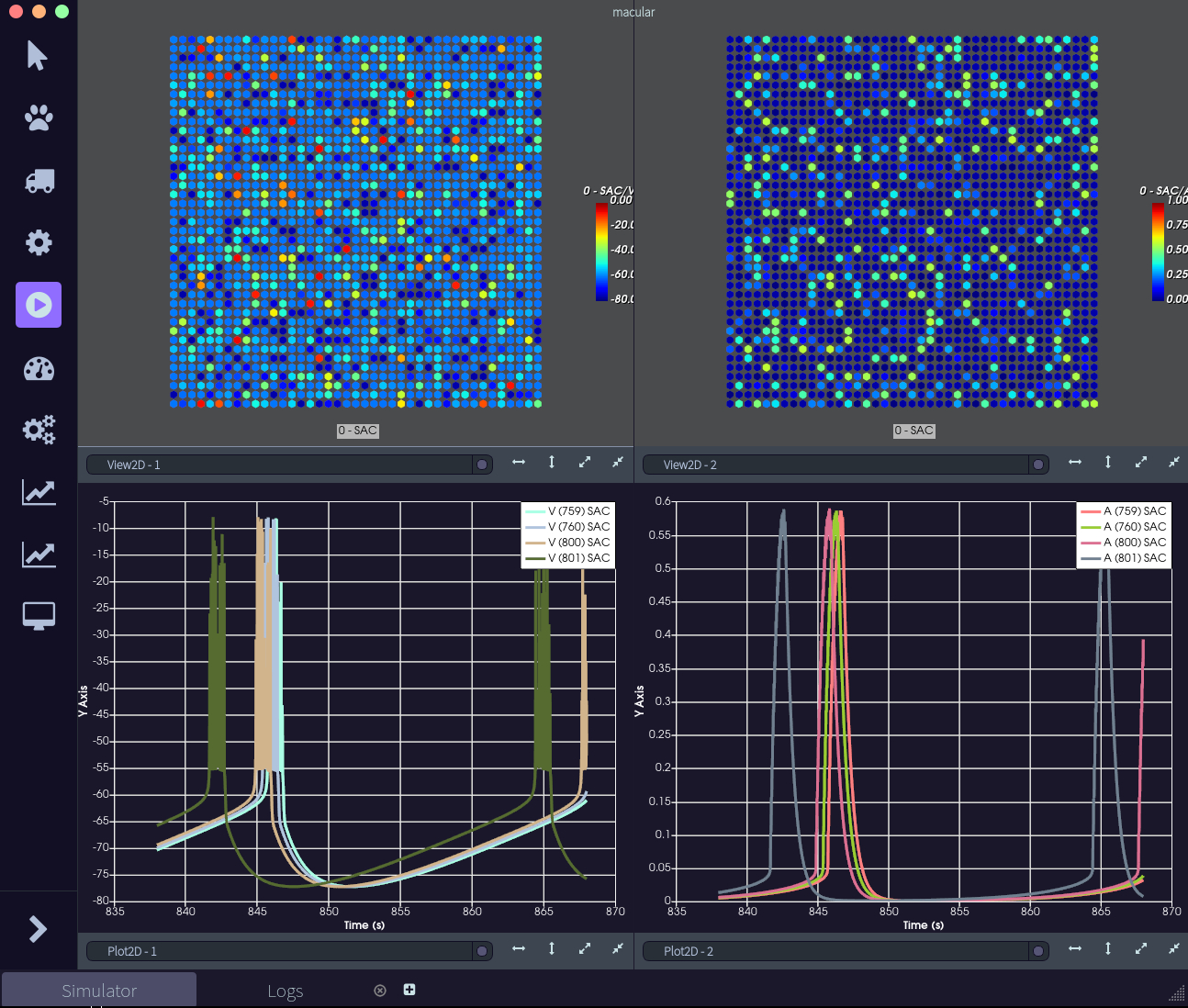}
\end{center}
\caption{\textbf{The retinal waves scenario.} \textbf{Top left.} Visualisation of the lattice evolution for the voltage of  Starburst Amacrine Cells (SACs).\textbf{Top right.} Visualisation of the lattice evolution for the acetylcholine concentration produced by SACs. \textbf{Bottom left.} Time evolution for the voltage of a few SACs selected by the user. \textbf{Bottom right.} Time evolution for the acetylcholine production of a few SACs.
\label{Fig:RW_Scenario}}
\end{figure*}

On \href{https://macular.gitlabpages.inria.fr/macular/user_doc/Macular/main.html#examples}{this page} we detail how to create the graph and how to run the simulation and visualise the results. A view of this simulation is shown in Fig. \ref{Fig:RW_Scenario}.

\sSec{A retino-cortical model}{Cortex} 

\begin{figure*}
\begin{center}
\includegraphics[width=\textwidth,height=0.5\textheight]{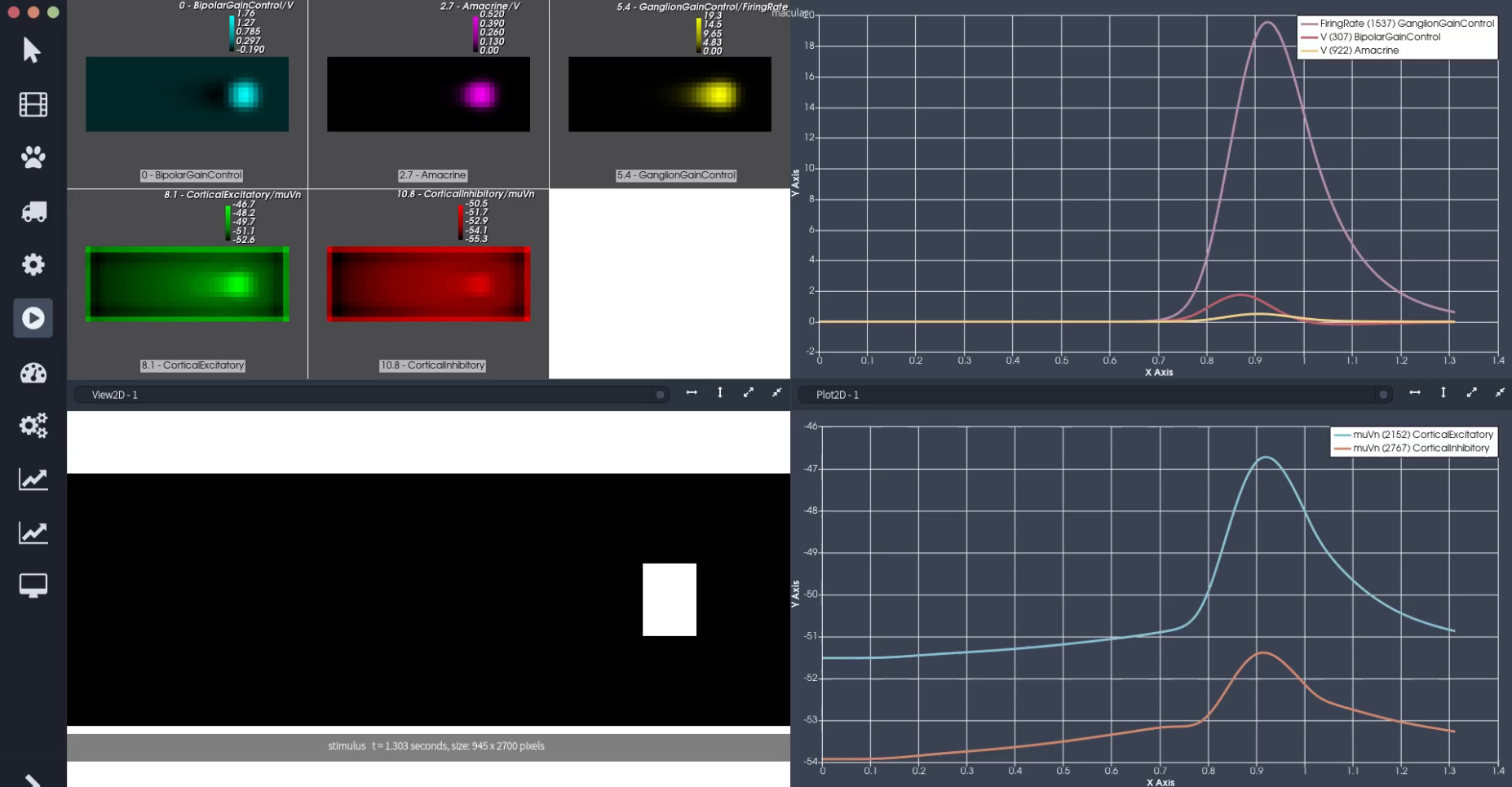}
\end{center}
\caption{\textbf{The retino-cortical scenario.} \label{Fig:ExampleSimulationRetinoCortical} The upper left panel shows the heatmap of the 5 cell types. Bipolar cells with gain control appear in blue, amacrine in magenta, ganglion cells with gain control in yellow, excitatory population of cortical columns in green and inhibitory population  of cortical columns in red. On the panel below, left, one sees the video of the  stimulus, a white bar moving. Right panels are plots of Cells activity. The upper one displays retinal outputs : bipolar voltage (red), amacrine voltage (yellow) and ganglion cells firing rate (pink). The bottom panel displays cortical output : excitatory (blue) and inhibitory (orange) mean voltage.}
\end{figure*}

In this second example, we consider a model of the retino-cortical associations featuring the joint evolution of the retina and V1 under visual stimuli. A view of the corresponding Macular simulation is shown in Fig. \ref{Fig:MacularGraph}
and Fig. \ref{Fig:ExampleSimulationRetinoCortical}.

The retina model is composed of $3$ layers: Bipolar cells with gain control (BCs) receiving an input from the OPL, Amacrine cells (ACs) providing lateral inhibition and retinal Ganglion cells (RGCs) receiving excitation from BCs and inhibition from ACs \citep{souihel-cessac:21,cessac:22,kartsaki-hilgen-etal:24,emonet-souihel-etal:25}. Moreover, BCs and ACs are mutually connected. BCs excite ACs, ACs inhibit BCs. It is possible to enable or disable the various features of the model (lateral connectivity and gain control) by adjusting its parameters.
The cortical model of V1 features the joint evolution of two populations of cortical column, one excitatory, the other inhibitory, coupled via delayed lateral connectivity depending of a conduction velocity, and evolving via dynamic mean field equations with an input coming from the retina/thalamus. This model has  has been proposed in \citep{elboustani-destexhe:09,benvenuti-chemla-etal:15,benvenuti-chemla-etal:20,zerlaut-chemla-etal:18}. We refer to these papers for the detail. This model and this Macular scenario  have been used in the papers \citep{emonet-cessac:25,emonet-souihel-etal:25}.
The firing rate of RGCs constitute the inputs of the cortical model. Thus, there is no thalamus in this example. 

How to create the corresponding Graph and generate the simulation is described in detail \href{https://macular.gitlabpages.inria.fr/macular/user_doc/Macular/main.html#examples}{there}. Here, we just show the result of a simulation obtained by loading a scenario available with the Macular release. In this scenario illustrated by Figure \ref{Fig:ExampleSimulationRetinoCortical}, the only feature activated is the lateral connectivity between bipolar and amacrine cells.
The corresponding Macular session and graph have been placed in the "macular/examples" repository : "Scenario2\_RetinoCortical.json" for the graph and "Scenario2\_RetinoCortical\_layout.json" for the session. 

\sSec{Creating a new model}{MTE_Creation}


Here, we give an example of new Cell type and Synapses created with the MTE. The detailed procedure can be found \href{https://macular.gitlabpages.inria.fr/macular/user_doc/Macular/main.html#examples}{here}.
 This example corresponds to the Amari-Wilson-Cowan model \citep{amari:71,wilson-cowan:73} whose equations reads:
\begin{equation}\label{eq:AWC}
\frac{dV_i}{dt}\,=\,-\frac{V_i}{\tau} \,+\, \sum_{j=1}^N J_{ij} f(V_j)    \,+\, H_{{ext}_i}(t), \quad j=1 \dots N,
\end{equation}
where $V_i$ is the voltage of Cell $i$ in mV, $\tau$ a characteristic integration time, $J_{ij}$ a synaptic weight (in mV/s),
$H_{{ext}_i}(t)$ the OPL input (it has the dimension  $mV s^{-1}$). The function $f$ is a sigmoid function of the form:
\begin{equation}\label{eq:sigmoid}
f(x)\,=\,\frac{1}{2}\,\bra{
1\,+\,erf\pare{
\frac{g\,x}{\sqrt{2}}
}
},
\end{equation}
$g$ being a positive parameter called "sigmoid gain" (in $mV^{-1}$). This model has been widely studied in the literature, especially in the case where the $J_{ij}$ are random independent variables, with no external current. In this case, dynamics is chaotic for a sufficiently high gain $g$ \citep{sompolinsky-crisanti-etal:88,cessac:19}. Here, we consider the case where the $J_{ij}$'s corresponds to Gaussian pooling (see section  \ref{Sec:Graph}, eq. \eqref{eq:GaussianPooling}) with excitatory and inhibitory synapses and with an OPL input. This does not really correspond to a realistic situation as \eqref{eq:AWC} would correspond to ganglion cells, the only spiking retinal cells,  that would receive a direct OPL input. This does not hold in the real retina.

\section{Comparison to other simulation software}\label{Sec:Alternatives}

Table \ref{Tab:Compare} shows a comparison between different retina simulation software.
We have been focusing here on software specialized to the retina. Thus, we didn't include more generalist sotware such as \href{https://www.comsol.com/}{COMSOL} (which also has a graphical interface), \href{https://fr.mathworks.com/products/matlab.html}{MATLAB}, or \href{https://www.neuron.yale.edu/neuron/}{NEURON}.



\begin{table*}[!ht]
\centerline{{\footnotesize
\begin{tabular}{|L{2cm}|C{12mm}|C{12mm}|C{12mm}|C{12mm}|C{12mm}|C{12mm}|C{12mm}|C{12mm}|C{12mm}|}
\cline{2-10}
\multicolumn{1}{c|}{ } & 
{\bfseries CR}&
{\bfseries RS}&
{\bfseries RT}&
{\bfseries CN}&
{\bfseries IS}&
{\bfseries P2P}&
{\bfseries RSt}&
{\bfseries VR}&
{\bfseries MA}\\
\hline 
OS 
& Linux
& All  
& Linux-Mac
& Linux 
& All 
& All 
& x 
& All 
& All \\ 
\hline 
Version 
& x 
& x 
& 1.7.57
& 0.6.4 
& x 
& 0.10.0 
& x 
& 2.2.3
& 1.5.2 \\ 
\hline 
Language 
& C++
& Matlab 
& C++
& Python 
& Matlab 
& Python 
& Flowlang 
& C++ 
& C++ \\ 
\hline 
Type 
& library
& toolbox 
& library
& toolbox 
& toolbox 
& library 
& library 
& standalone 
& standalone \\ 
\hline 
Dependences 
& \no 
& Matlab 
& \no 
& PyTorch 
& Matlab 
& \no 
& x 
& \no 
& \no \\ 
\hline 
Open source 
& \yes 
& \no * 
& \yes 
& \yes 
& \yes 
& \yes 
& \yes 
& \yes 
& \yes \\ 
\hline
P.K.R.
& \yes 
& \no 
& \yes 
& \yes 
& \no 
& \yes 
& \yes 
& \yes 
& \no \\ 
\hline
GUI 
& \no
& \no
& \no 
& \no 
& \no 
& \no 
& \no 
& \no 
& \yes \\ 
\hline
Videos as inputs
& \yes 
& \no 
& \no 
& \yes 
& \yes 
& \yes 
& \yes 
& \yes 
& \yes \\ 
\hline
OPL modelling
& S.T.K. 
& D.E. 
& x 
& S.T.K. 
& N.T. 
& S.T.K. 
& S.T.K. 
& S.T.K. 
& S.T.K. \\ 
\hline 
Shape of RF
& x 
& x 
& x 
& Any 
& x 
& N.A. 
& x 
& circular 
& circular \\ 
\hline
3D visualisation
& \no 
& \no 
& \no 
& \no 
& \no 
& \no 
& \no 
& \no 
& \yes \\ 
\hline
Scripting Interface 
& \no 
& \no
& \no 
& \no 
& \no 
& \no 
& \yes 
& \no 
& \yes \\ 
\hline 
Parameters tuning 
&  
& File 
& File 
& Automatic 
& File 
& File 
& x 
& File 
& Sliders \\ 
\hline 
Extended cells 
& \no 
& \no 
& \no 
& \no 
& \no 
& \no 
& \no 
& \no 
& \yes \\ 
\hline 
Thalamus, Cortex 
& \no 
& \no 
& \no 
& \yes  
& \yes  
& \no 
& \no 
& \no 
& \yes \\ 
\hline Cells recording 
& \yes 
& \yes
& \yes  
& \yes 
& \yes 
& \no 
& \no 
& \yes 
& \yes \\ 
\hline
I.W.O.L.  
& NEST
& Matlab 
& \no 
& \no 
& Matlab 
& \no 
& \no 
& \no 
& \no \\ 
\hline
S.P.U. 
& \no 
& \no 
& \no 
& \no 
& \no 
& \no 
& \no 
& \no 
& \yes \\ 
\hline
Prostheses 
& \no
& \no 
& \no 
& \no 
& \no 
& \yes 
& \yes 
& \no 
& \yes \\ 
\hline 
C.I.M.
& \no 
& \yes 
& \no 
& \no 
& \no 
& \no 
& \no 
& \no 
& \yes \\ 
\hline
\end{tabular}
}}
\captionsetup{width=1.1\linewidth}
\caption{\label{Tab:Compare}
Comparison between a selection of existing software and Macular. Abbreviations for software names  have been chosen for the presentation. Following software is discussed:
{\bfseries CR}: \hrefi{\sc COREM}{https://www.worldscientific.com/doi/abs/10.1142/S0129065716500301}~\citep{martinez-canada-morillas-etal:16}, 
{\bfseries RS}: \hrefi{\sc "Retina Simulator"}{https://ieeexplore.ieee.org/abstract/document/9073963}~\citep{baek-eshraghian-etal:20}, 
{\bfseries RT}: \hrefi{\sc RetSim}{https://retina.anatomy.upenn.edu/~rob/ncman6.html}. 
{\bfseries CN}: \hrefi{\sc CONVIS}{https://www.frontiersin.org/articles/10.3389/fninf.2018.
00009/full?report}~\citep{huth-masquelier-etal:18}, 
{\bfseries IS}: \hrefi{\sc ISETBio}{https://github.com/isetbio/isetbio/wiki}~\citep{cottaris-jiang-etal:19}, 
{\bfseries P2P}: \hrefi{\sc Pulse2Percept}{https://www.biorxiv.org/content/biorxiv/early/2017/07/10/148015.full.pdf}~\citep{beyeler-boynton-etal:17}, 
{\bfseries RSt}: \hrefi{\sc RetinaStudio}{https://www.sciencedirect.com/science/article/pii/S0925231212007850}, 
{\bfseries VR}: \hrefi{\sc Virtual Retina}{https://team.inria.fr/biovision/virtualretina/}~\citep{wohrer-kornprobst-etal:09}, and 
{\bfseries MA}: \hrefi{\sc Macular }{https://team.inria.fr/biovision/macular-software/}. 
Note that features selected in this table have essentially been chosen according to what Macular\ does.  It is not an exhaustive list and information applies to the time of writing. Software we mention can have additional features not commented herein.
"All" in the row "OS (operating system)" means Linux, Mac and Windows.
 Other abbreviations used in this table are:
D.E. Differential Equations.
P.K.R: Programming Knowledge Required.
I.W.O.L: Interface With Other Languages.
S.P.U: Select Physical Units.
C.I.M: Choice of Integration Methods.
S.T.K. Spatio-temporal Kernels.
N.T. Nonlinear Transformations.
A \yes means "yes".
A white cell means "no".
A x in a column means that we haven't been able to find the information.
"Extended cells", signifies the Cell concept of Macular. * means that the referred Git page is not accessible.
}
\end{table*}

\section{Discussion}\label{Sec:Conclusion}

Macular was designed with the idea of maximising its accessibility, usability and development. That is why it is distributed as free software. It is also why we have designed an interface that allows non-programmers to use it, enabling them not only to simulate existing scenarios, but also to design new ones by creating new types of cells and synapses. 

We would now like Macular to evolve freely, according to the communities that might use it. With this in mind, there could be several useful developments:

\begin{itemize}
\item \textbf{Multiple cell classes.} 
Macular allows to simulate different cell classes simultaneously (e.g. BCs, ACs, RGCs)  and different types within each class. So we can simulate  ON and OFF BCs at the same time. However, the input to BCs comes from VR that emulates the OPL and that response can be either ON or OFF. We are currently working on an extended worker allowing to feature ON and OFF OPL responses simultaneously.
\item \textbf{Point neurons.} In the current release multi-compartment models of neurons are not allowed. It is however possible to upgrade Macular to have spatially extended neurons, although this is not planned by our group.
\item \textbf{Interfaces with other simulators.} As we have shown, Macular can be used to produce a model of the V1 cortex receiving realistic retinal input (see \cite{emonet-souihel-etal:25}). To our knowledge, this is the only example of its kind. It would be interesting to extend the integration to other cortical areas by interfacing Macular with other simulators such as \href{https://www.thevirtualbrain.org/tvb/zwei/home}{TheVirtualBrain} \citep{sanz-leon-knock-etal:13} or \href{https://www.nest-simulator.org/}{NEST} \citep{gewaltig-diesmann-etal:01}.
\item \textbf{Optimisation of computation time.} It would be beneficial to port Macular to parallel architectures or GPUs, which would allow for larger-scale or real-time simulations (in the spirit of \citep{baek-eshraghian-etal:20}).
\item \textbf{Generalised receptive fields.} The method we use to calculate the convolution of receptive fields with stimuli, imported from VirtualRetina, does not allow for asymmetric kernels, for example sensitive to direction or orientation, unlike simulators such as Convis \citep{huth-masquelier-etal:18}. It would be useful to extend the kernels of the receptive fields to a more general form, at the cost of slower calculations \citep{huth-masquelier-etal:18}.
\end{itemize}

We hope that Macular and its extensions will enable a new type of simulation in neuroscience, allowing for integrated models of the visual system with realistic sensory inputs, i.e. dynamic and multi-scale in space and time, e.g. films presenting visual scenes from the external world.

\section*{Conflict of Interest Statement}

The authors declare that the research was conducted in the absence of any commercial or financial relationships that could be construed as a potential conflict of interest.

\section*{Author Contributions}

B.C. has supervised the Macular project, participated to its elaboration, tests, and design, contributed to the use cases, and to the online documentation. He wrote the paper.
T.K., C.L.B., J.L.S, and J.W. contributed to Macular design and conception as well as software development.
E.D., N.N.  contributed to Macular design and conception,  software development and to the online documentation.
J.E., E.K. and S.S. contributed to Macular design and conception, to the software development and tests,  and to the use cases. 

\section*{Funding}
This work was funded by the Inria AMDT. It was supported by the Leverhulme Trust (RPG-2016-315) funding Evgenia Kartsaki's PhD, the National Research Agency (ANR), in the project “Trajectory”, https://anr.fr/Project-ANR-15-CE37-0011,
 funding Selma Souihel’s PhD; the ANR too in the project "Shooting Star-15755" https://anr.fr/Projet-ANR-20-CE37-0018, funding Jérôme Emonet’s PhD, and finally
by the interdisciplinary Institute for Modelling in Neuroscience and Cognition (NeuroMod http://univ-cotedazur.fr/en/idex/projet-structurant/neuromod) of the Université Côte d'Azur.

\section*{Acknowledgments}
We thank Téva Andreoletti, Ghada Balhoul, Eléonore Birgy, Tristan Cabel, Simone Ebert, Pierre Fernique, Sebastián Gallardo, Jonathan Levy, Andres Navarro, Alex Ye, Carlos  Zubiaga, for their help in developing or testing Macular.


\section*{Data Availability Statement}
The source code of this software can be found in the Git repository \url{https://gitlab.inria.fr/macular/macular}.


\bibliographystyle{abbrv} 
\bibliography{odyssee}


\subsection{Tables}

\end{document}